\documentclass[prl,twocolumn,showpacs,preprintnumbers,amsmath,amssymb, superscriptaddress]{revtex4}
\usepackage{graphicx}
\usepackage{dcolumn}
\usepackage{bm}

\begin{document}


\title{
Orbitally-driven Peierls state in spinels
}
\author{D.~I.~Khomskii} 
\affiliation{II Physikalisches Institut, Universit\"at zu K\"oln, 50937
K\"oln, Germany}
\affiliation{Groningen University, Nijenborgh 4, 9747 AG Groningen, The Netherlands}

\author{T.~Mizokawa}
\affiliation{
Department of Physics and Department of Complexity Science and Engineering, 
University of Tokyo, 5-1-5 Kashiwanoha, Chiba 277-8581, Japan
}

\date{\today}

\begin{abstract}
We consider the superstructures, which can be formed in spinels containing
on B-sites the transition-metal ions with partially filled $t_{2g}$ levels.
We show that, when such systems are close to itinerant state (e.g. have an
insulator-metal transition), there may appear in them an orbitally-driven 
Peierls state. We explain by this mechanism the very unusual superstructures
observed in CuIr$_2$S$_4$ (octamers) \cite{Radaelli} and MgTi$_2$O$_4$ 
(chiral superstructures) \cite{Schmidt} and suggest that similar 
phenomenon should be observed in NaTiO$_2$ and possibly 
in some other systems.
\end{abstract}


\maketitle

Systems with complicated lattices, such as pyrochlore-type lattice of 
corner-sharing tetrahedra in spinels, often display
complicated types of ordering. The best known example is magnetite
Fe$_3$O$_4$, which apparently has charge ordering below Verwey transition
at $T_V$ = 120 K, the detailed type of ordering still being a matter of
hot debate \cite{Fe3O4}. Recently two other examples of exotic types of
ordering were observed by Radaelli and coworkers: octamer ordering
in CuIr$_2$S$_4$ \cite{Radaelli} and helical (or chiral) distortion
in MgTi$_2$O$_4$ \cite{Schmidt}. In all three cases, superstructures
appear in the insulating phase below metal-insulator transition;
in CuIr$_2$S$_4$ and MgTi$_2$O$_4$, it is accompanied by the formation
of nonmagnetic state (spin singlet on a dimer). The origin of these 
strange and beautiful structures was not clarified in \cite{Radaelli,Schmidt}

Below we suggest a very simple explanation of the superstructures
in CuIr$_2$S$_4$ and MgTi$_2$O$_4$, using the concept of 
{\em orbitally-driven Peierls state}. We show that the specific
features of orbitals involved, on the pyrochlore lattice of B-sites
of spinels, lead to the formation of essentially one-dimensional
bands and, with proper filling, lead to the Peierls-like effect,
which in these particular cases has a form of tetramerization
along certain directions. Viewed from different perspective,
the resulting states give just the octamer ordering of 
Ref. \cite{Radaelli} and chiral ordering of Ref. \cite{Schmidt}.
Using the same concept, we predict also that similar phenomenon
should exist in some other materials, e.g., in NaTiO$_2$
and possibly in V spinels.

\begin{figure}
\includegraphics[width=8.0cm]{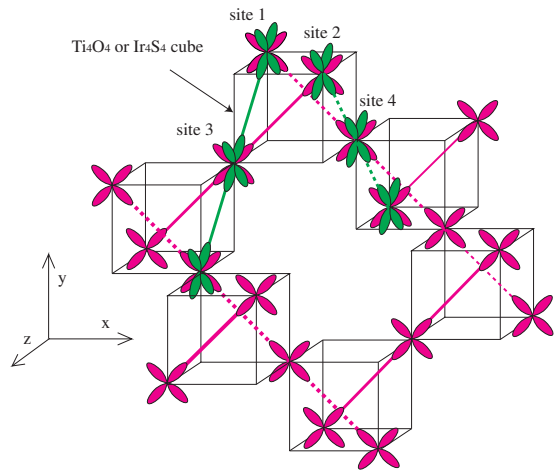}
\caption{
Orbital dependent hopping in the pyrochlore lattice of spinels.
For an electron in the $xy$ orbital of site 1, the dominant hopping
is along the (1, -1, 0) direction in the $xy$ plane (to the site 4). 
On the other hand,for an electron in the $xy$ orbital of site 2, the dominant hopping
is along the (1, 1, 0) direction in the $xy$ plane (to the site 3). The (1, 1, 0)
chains shown by the red solid line never crosses the (1, -1, 0) chains
shown by the red broken line. For electrons with the $yz$ orbitals
of site 1 and 2, the dominant hopping is along the (0, 1, -1) 
and (0, 1, 1) directions (to the sites 3 and 4, respectively).
}
\label{cube}
\end{figure}

In both CuIr$_2$S$_4$ and MgTi$_2$O$_4$, the transition-metal (TM) 
ions Ir and Ti are located in almost regular S$_6$ and O$_6$ octahedra,
so that the dominant crystal field (CF) splits the $d$-levels into
the $t_{2g}$ triplet and $e_g$ doublet. In both cases, only $t_{2g}$
levels are occupied, by one electron in Ti$^{3+}$ and by five or six
electrons in Ir$^{4+}$ and Ir$^{3+}$ respectively
(average Ir valence is +3.5).
The shape of these orbitals is such that the strongest overlap 
in the elementary Ir$_4$S$_4$ or Ti$_4$O$_4$ cubes is between
the same orbitals along the particular direction, e.g., 
$xy$ with $xy$-orbitals for the TM pair in the $xy$-plane,
or $yz$ with $yz$-orbitals for the pair in the $yz$-plane,
see Fig. \ref{cube}. Thus the dominant hopping will be
along the straight TM-TM chains, which would lead to
formation of one-dimensional bands with the dispersion
$E(k_\alpha)=-2tcosk_{\alpha}$ where $\alpha$ denotes 
the direction along corresponding chain. 
This orbital-dependent hopping will play crucial role 
in our explanation.

\begin{figure}
\includegraphics[width=8.0cm]{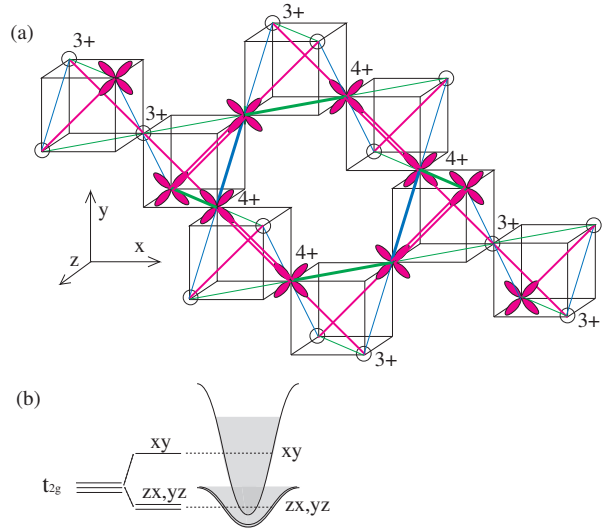}
\caption{(a) Charge and orbital ordering in CuIr$_2$S$_4$. 
Octamer is shown by thick lines, short singlet bonds --- by double lines.
(b) Schematic electronic structure of CuIr$_2$S$_4$.}
\label{CuIr2S4}
\end{figure}

We start by discussing the case of CuIr$_2$S$_4$. 
In this system, below the metal-insulator transition temperature $T_{MI}$, 
there appears the net tetragonal distortion
(elongation, $c/a$ = 1.03 \cite{CuIr2S4}), and besides that,
the complicated octamer structure appears \cite{Radaelli}:
Ir$^{3+}$ and Ir$^{4+}$ order in octamers, and in the Ir$^{4+}$
octamers there appears also an alternation of short and long bonds,
see Fig. 2 in \cite{Radaelli}. This beautiful structure seems extremely
unusual. However the situation is much simpler if one looks 
at what happens in the {\em straight Ir chains}: one immediately notices
that in five out of six such chains there appears a tetramerization
--- an alternation Ir$^{3+}$/Ir$^{3+}$/Ir$^{4+}$/Ir$^{4+}$/...,
and in one of them --- corresponding dimerization, 
see Fig. \ref{CuIr2S4}(a).

One can naturally explain this pattern if one looks 
at the electronic structure of this compound, schematically shown
in Fig. \ref{CuIr2S4}(b). Due to tetragonal elongation, there will
appear a CF splitting of $t_{2g}$ levels, and besides (which is 
probably more important), the $xy$ band will become broader.
With the 5.5 electrons (or 0.5 hole) per Ir in these levels,
the lowest two bands will be fully occupied, and the upper
$xy$ one-dimensional band will be 3/4-filled. 
Thus we can expect a Peierls or charge density wave (CDW) transition,
accompanied by tetramerization in the $xy$ chains [along 
the (1, 1, 0) and (1, -1, 0) directions], with holes 
in the $xy$ orbitals, as shown in Fig. \ref{CuIr2S4}(a).
As is seen from this figure, the resulting state exactly corresponds
to the one found in \cite{Radaelli}: Ir$^{3+}$ and Ir$^{4+}$ form
octamers. Besides, the Ir$^{4+}$ pairs in the $xy$ chains
have orbitals directed towards one another, thus these pairs will
form spin singlets. When we release the lattice, corresponding bonds
will become shorter, again consistent with the structure of \cite{Radaelli}.
Thus the explanation of this apparently very complicated structure
becomes extremely simple and natural if we look at it from the point
view of straight Ir chains, which, for this orbital occupation,
form natural building block in spinels.

\begin{figure}
\includegraphics[width=8.0cm]{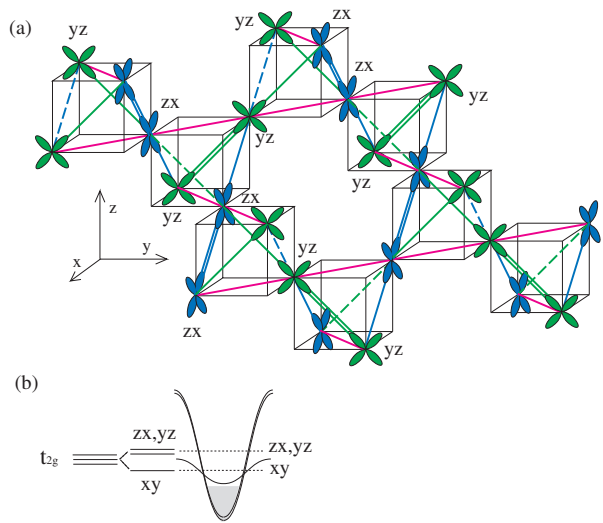}
\caption{
(a) Orbital ordering in MgTi$_2$O$_4$. 
Short singlet bonds are shown by double, intermediate --- single, 
and long --- dashed lines. $yz$ orbitals are shown by green
and $zx$ orbitals --- by blue.
(b) Schematic electronic structure of MgTi$_2$O$_4$. Note different orientation of coordinate axes as compared with Figs.1,2 }
\label{MgTi2O4}
\end{figure}

The same idea explains also the chiral superstructures
observed in MgTi$_2$O$_4$ \cite{Schmidt}.
Below $T_{MI}$ at 260 K, a tetragonal distortion (here contraction)
appears also in this system, together with the inequivalent bonds,
so that, if one connects short and long bonds, they form spirals
along the $c$- or $z$-direction, which may be both left- and
right-moving. Apparently, on the short bonds, Ti-Ti pairs form
spin singlets which is rather typical for $d^1$ configurations.
This naturally explains the drop of magnetic susceptibility
below $T_{MI}$ \cite{Ueda}.
And again this superstructure, the origin of which looks very puzzling,
can be explained very easily if one looks at the situation in the
{\em straight Ti chains}. One immediately notices that in all chains
running in the (0, 1, 1), (0, 1, -1), (1, 0, 1), and (1, 0, -1)
directions (lying in the $zx$- and $yz$-planes)
one has the {\em tetramerization}: an alternation of 
short/intermediate/long/ intermediate/... bonds. 
This structure can be naturally explained 
if we look at the electronic structure of this system, Fig. \ref{MgTi2O4}(b). 
In the high temperature phase, Ti$^{3+}$ ions 
have one electron in the triply-degenerate $t_{2g}$ level,
which in the tight-binding scheme would give three one-dimensional
degenerate bands (we neglect here small trigonal splitting).
One can reduce the band energy by tetragonal distortion
--- the effect similar to the band Jahn-Teller effect
invoked by Labbe and Friedel to explain the cubic-tetragonal
transition in A15 compounds (V$_3$Si, Nb$_3$Sn) \cite{A15}.
The tetragonal compression observed in MgTi$_2$O$_4$ would
increase the bandwidths of the $zx$ and $yz$ bands
and decrease that of the $xy$ band, opposite to the case
of CuIr$_2$S$_4$. With one electron per Ti, the lowest doubly
degenerate $zx$ and $yz$ bands will be {\em 1/4-filled}.
This would lead to the usual instability and to the formation
of superstructure with the wave vector $Q_{\alpha}=\pi/2$, i.e.
to a tetramerization in the $zx$- and $yz$-directions,
in accordance with the experiment. In contrast to the usual
Peierls transition, we even do not have here to move ions:
one can get corresponding superstructure by changing orbital
occupation of respective ions (we may call it a ODW ---
orbital density wave).

Such ODW will be stabilized by the electron repulsion. 
In the strong-coupling limit we would get in each of 
these chains the orbital ordering of the type 
$zx$-$zx$-$yz$-$yz$-$zx$-$zx$-..., see Fig. \ref{MgTi2O4} 
(this ordering satisfies the "Anderson rule":
there will be two orbitals of each type in each Ti tetrahedron). 
Of course the lattice would then follow: 
strong overlap of the orbitals in the pair with the lobes
along corresponding chains would lead to the 
alternation of strong (short)-intermediate-weak(long)-intermediate
bonds in each chain, with the spin-singlet states at short bonds. 
If now one connects short with long bonds, one gets the helical
pattern stressed in \cite{Schmidt}.
Thus one can visualize the structure observed in \cite{Schmidt} 
both as a bond alternation and as a site-centered orbital ordering. 
Here we considered tetragonal transition 
(band Jahn-Teller effect) and orbitally-driven Peierls distortion 
separately, but in reality they occur simultaneously, so that 
it is the total energy gain which stabilizes this structure.

The situation in MgTi$_2$O$_4$ is somewhat more complicated 
than that in CuIr$_2$S$_4$: whereas in the latter 
both the local CF splittings and the change of the bandwidths 
combine to make only $xy$ band partially occupied, 
in MgTi$_2$O$_4$ these two effects oppose one another. 
Apparently the band Jahn-Teller effect here dominates 
leading to the tetragonal elongation instead of compression
which would be more favourable for localized electrons.
One can make crude estimate, which shows that one gains
much more kinetic energy with this filling 
(almost 0.5$t$, where $t$ is the nearest-neighbour hopping)
when one occupies doubly-degenerate bands close to the bottom 
(which happens at distortion with $c/a<1$) than 
if one would have an opposite distortion and would 
occupy nondegenerate $xy$ band. But the most direct 
confirmation of our picture comes from the band-structure
calculations presented in Ref.\cite{Schmidt}: 
according to them below $T_c$ in MgTi$_2$O$_4$ 
indeed only the $zx$ and $yz$ bands are occupied.

One can argue that similar phenomenon should exist 
in other materials, e.g. in NaTiO$_2$ with 
the triangular two-dimensional lattice of Ti$^{3+}$ ions. 
This material also has a metal-insulator transition 
with structural distortion at 250K, which leads to 
a rather strong elongation of TiO$_6$ octahedra 
along one of the local pseudocubic axes \cite{NaTiO2} 
(which, in combination with the original rhombohedral 
structure, makes the system monoclinic). 
The resulting situation may then become similar 
to that in MgTi$_2$O$_4$, and one can expect 
the formation of similar superstructure 
with the alternation of $zx$-$zx$-$yz$-$yz$-$zx$-$zx$-... orbitals 
along two sets of Ti chains running along the distortion axis, 
with the formation of corresponding spin-singlet states. 
The drop of susceptibility was indeed observed in NaTiO$_2$ 
below $T_c$ \cite{NaTiO2}, but the predicted superstructure was not yet seen.
 
The situation may be also similar in V spinels $M$V$_2$O$_4$, 
$M$ = Mg,Zn,Cd \cite{Mamiya, Lee}. 
All of them have cubic-tetragonal transitions with $c/a<1$, 
which lead to the splitting of $t_{2g}$ levels with the low-lying 
$xy$ singlet and higher $zx$, $yz$ doublet. With two electrons per V ion, 
we would have the situation resembling that of Ti : one electron 
would occupy the $xy$ orbital, but the second one would be 
in the doubly-degenerate ($zx$, $yz$) orbitals or bands 
(note that this distortion  ($c/a<1$) is again "anti-Jahn-Teller": 
Jahn-Teller effect for localized electrons would rather lead to 
the opposite distortion, with both electrons in the lower doublet).
Consequently, one may expect the formation of the  
superstructure similar to that discussed above. Recent results \cite{Lee} 
confirm the presence of one-dimensional spin chains made 
by the $xy$-electrons, and show some indirect evidences 
of superstructure in the $c$- or $z$-direction, which, however, 
the authors interpret as dimerization (alternation of planes 
with $zx$ and $yz$ orbitals, see also \cite{Tsunetsugu}). 
More detailed structural studies should clarify which structure 
is actually realized in these systems. 

And finally, the same effect may be relevant in magnetite, 
at least partially. The most recent structure proposed for Fe$_3$O$_4$ 
\cite{Wright} (shown in Fig.7 of \cite{Wright}) has certain similarities 
with the structures discussed above: there is also a tetramerization 
in $zx$- and $yz$-chains, but of somewhat different type than 
that in CuIr$_2$S$_4$: 
in two out of four such chains we have the same pattern, 
Fe$^{2+}$/Fe$^{2+}$/Fe$^{3+}$/Fe$^{3+}$/..., 
but in two others the tetramerization is of different type: 
in one 2+/2+/2+/3+/2+/2+/2+/3+/... and in another 3+/3+/3+/2+/3+/3+/3+/2+/... 
According to LDA+$U$ calculations \cite{Leonov}, there appears 
also an orbital ordering in this structure, such that again 
only two types of orbitals $zx$ and $yz$ (or their combinations) 
are occupied by the extra $t_{2g}$ electron of Fe$^{2+}$. 
It is not yet clear to us whether the effects studied above 
play a role in this ordering in magnetite, but the analogy 
is rather suggestive.

Actually, the situation with orbitally-driven Peierls state 
leading to spin singlets is not unique for spinels or systems 
with triangular lattices like in NaTiO$_2$ \cite{Horsch}.
Such phenomenon was also observed in pyroxene NaTiSi$_2$O$_6$ \cite{Isobe}. 
One can also argue that the same effect --- 
orbitally-driven singlet pair formation --- 
occurs also in the recently synthesized La$_4$Ru$_2$O$_{10}$ \cite{Khalifah}. 
The explanation given in Ref. \cite{Khalifah} 
(transition to a zero-spin Ru$^{4+}$ state) 
seems rather unlikely, as it would require CF splitting 
of Ru $t_{2g}$ levels exceeding the on-site Hund's rule energy, 
which seems rather improbable. Rather, one can suggest
that the nonmagnetic ground state of La$_4$Ru$_2$O$_{10}$ is
also caused by the orbitally-driven spin singlets. 
Recent neutron scattering results on single crystalline La$_4$Ru$_2$O$_{10}$
\cite{Osborn} may indeed agree with this interpretation.
Thus there appears more and more systems in which an orbital
ordering leads to a singlet Peierls state.

Summarizing, we argue that the complicated superstructures 
observed recently in some spinels (CuIr$_2$S$_4$ and MgTi$_2$O$_4$) 
are naturally explained as an orbitally-driven Peierls distortion, 
caused by the proximity to an itinerant state.
We predict similar phenomenon in NaTiO$_2$, and possibly 
in some other systems.

We are grateful to P.~Radaelli, S.-H.~Lee and R.~Osborn for useful discussions.

\end{document}